\documentclass[fleqn]{article}
\usepackage[utf8]{inputenc}
\usepackage[ukrainian]{babel}
\usepackage[T1]{fontenc}
\usepackage[dvips,final]{graphicx}

\usepackage{amsmath}
\usepackage{amsfonts}
\usepackage{mathrsfs}

\newtheorem{theorem}{Теорема}

\newtheorem{definition}{Означення}

\def\la{\lambda }
\def\lt{\lambda  t}

\def\ltop#1{\overline{l_{#1}}}
\def\lbot#1{\underline{l_{#1}}}

\begin{document}%%%%%%%%%%%%%%%%%%%%%%%%%%%%%%%%%%%%%%%%%%%%%%%%%%%%%%%%%%%%%%%%%%%%

\date{6 березня 2015 р.}

\author{О.К.Відибіда\\
Інститут теоретичної фізики ім. М.М.Боголюбова,\\
        вул Метрологічна, 14-Б, 03680 Київ, Україна\\ email: vidybida@bitp.kiev.ua\\
http://www.bitp.kiev.ua/pers/vidybida\\т.4243037, 5213467, 0663398203}

\title{Вихідний потік інтегруючого нейрона з втратами. I. Розподіл вихідних міжімпульсних інтервалів}

\maketitle

\thispagestyle{empty}

\vfill
\small
$^*$Перші три зноски --- для полегшення перекладу на англійську і в українському варіанті мають бути опущені.

\newpage

\setcounter{page}{1}

\noindent
{\bf О.К.Відибіда} (Інститут теоретичної фізики ім. М.М.Боголюбова, Київ)\bigskip

\thispagestyle{empty}

\noindent
{\large\bf ВИХІДНИЙ ПОТІК ІНТЕГРУЮЧОГО НЕЙРОНА З ВТРАТАМИ. I. Розподіл вихідних міжімпульсних інтервалів}\bigskip

\noindent
{\large\bf OUTPUT STREAM OF LEAKY INTEGRATE AND FIRE NEURON. I. Distribution of output interspike intervals}\bigskip

\noindent
Обчислено в явному вигляді густину розподілу ймовірності вихідних
міжімпульсних інтервалів для інтегруючого нейрона з втратами стимульованого
процесом Пуассона. Не застосовується диффузійне наближення. Одержаний розподіл
характеризується бімодальністю при певних значеннях параметрів.
\bigskip

\noindent
Probability density function of output interspike 
intervals is found in exact form for leaky integrate and fire neuron
stimulated with Poisson stream. The diffusion approximation is not used.
The distribution found can be bimodal for some values of parameters.
\bigskip

\section{Вступ}

Інтегруючий нейрон з втратами, \cite{Stein1967}, це найбільш вживана математична модель нейрона в 
теоретичних нейронауках, що пояснюється її відносною простотою, див. Розділ \ref{OLIF}.
Разом з тим, для цієї моделі задача знаходження ймовірносного розподілу довжин 
вихідних міжімпульсних інтервалів при стимуляції потоком Пуассона не розв’язана.
Частково це пояснюється тим, що імпульс, одержаний нейроном розпадається за 
показниковим законом і його залишок може бути присутній в нейроні протягом часу, 
який обмежується тільки моментом наступного 
пострілу\footnote{В англомовних роботах вживається термін ``firing moment''. Це такий
момент, коли досягається порогове збудження і нейрон посилає вихідний імпульс 
після чого повертається до стану спокою.}.

Численні результати що до опису вихідного потоку одержано при застосуванні
диффузійного наближення, див. огляд \cite{Sacerdote2013}.  В диффузійному наближенні
часовий хід мембранного потенціалу (величини збудження) описується як процесс
Орнштейна-Уленбека, \cite{Ricciardi}. При цьому обчислюється середнє міжімпульсних
вихідних інтервалів і їхня дисперсія. Ймовірносний розподіл довжин міжімпульсних
інтервалів описується здебільшого наближено в силу складності обчислення розподілу
часів першого досягнення порогу для процесу Орнштейна-Уленбека, див. \cite{Alili2005}.

Застосування диффузійного наближення є доцільним коли для досягнення порогового збудження
потрібно багато вхідних імпульсів, які надходять через короткі проміжки часу. Така ситуація
має місце для деяких нейронів, \cite{Andersen1990}. Разом з тим, існують нейрони, для збудження яких потрібна невелика кількість вхідних імпульсів, починаючи з двох, 
\cite{Miles1990,Gulyas1993}. Для таких нейронів диффузійне наближення не буде обгрунтованим.

В цій роботі обчислюється густина ймовірності розподілу довжин вихідних міжімпульсних
інтервалів без застосування диффузійного наближення. Можливість такого обчислення
забезпечується накладеними обмеженнями  на співвідношення між величинами
вхідного імпульсу і порогового збудження, див. (\ref{cond}), нижче. Значення густини ймовірності для певної
довжини міжімпульсного інтервалу $t$ одержується в вигляді скінченної сумми кратних
інтегралів.

\section{Постановка задачі}

\subsection{Означення інтегруючого нейрона з втратами}
\label{OLIF}

Інтегруючий нейрон з втратами (ІНВ)\footnote{В англомовній літературі --- leaky integrate and fire (LIF).}
характеризується трьома позитивними константами:
\begin{enumerate}
	\item $\tau$ --- час релаксації,
	\item $V_0$ --- поріг збудження,
	\item $h$ --- величина вхідного імпульсу.
\end{enumerate}
Відносно $h$ і $V_0$ ми робимо наступне припущення:
\begin{equation}
\label{cond}
	0<h< V_0 < 2h.
\end{equation}

В будь-який момент $l\in[0;\infty[$ стан ІНВ характеризується невід’ємним 
дійсним числом $V(l)$,
яке інтерпретується як відхилення трансмембранної різниці потенціалів від 
стану спокою в бік деполяризації, або іншими словами 
величина збудження. Тут вважається, що в стані спокою $V=0$,
а деполяризації/збудженню відповідає позитивне значення $V$.

Наявність втрат означає, що за відсутності 
зовнішніх стимулів величина $V(l)$ експоненційно зменшується:
\begin{equation}
\label{tau}
	V(l+s)=V(l) e^{-\frac{s}{\tau}},
	\qquad s>0.
\end{equation}

Вхідні стимули --- це вхідні імпульси.
Одержання вхідного імпульсу в момент $l$ 
підвищує $V(l)$ на величину $h$:
\begin{equation}
\label{h}
	V(l) \rightarrow V(l) + h.
\end{equation}

Нейрон характеризується пороговим значенням збудження $V_0$.
Останнє означає, що як тільки виконано умову
$%\begin{equation}
\label{V0}\nonumber
	V(l) > V_0,
$%\end{equation}
\ ІНВ стріляє, 
т.т. генерує вихідний імпульс, який в біологічній літературі називається спайк,
і переходить в стан спокою, $V(l)=0$.

З (\ref{tau}) і (\ref{h}) випливає, що ІНВ може 
згенерувати вихідний імпульс тільки в момент одержання вхідного.
Це узгоджується з поведінкою реальних нейронів з певною мірою точності.
Умова (\ref{cond}) означає, що одного вхідного імпульсу,
застосованого до ІНВ в стані спокою,
не досить для пострілу, але вже два вхідних імпульси,
отримані за короткий проміжок часу, можуть збудити ІНВ достатньо для 
генерації вихідного імпульсу\footnote{В англомовних статтях замість
``збудити нейрон достатньо для генерації вихідного імпульсу''
 вживається термін ``to trigger neuron''.}.
Ми прийняли обмеження (\ref{cond}),
оскільки запропонований нижче математичний метод виглядає прозоріше при виконанні (\ref{cond}).
Разом з тим, дієвість запропонованого методу при виконанні замість (\ref{cond}) умови 
$	2h\le V_0 < 3h$
і інших аналогічних умов не виключається.

\subsection{Опис процесу Пуассона}

Вважається, що ІНВ одержує вхідні імпульси від стохастичного
процесу Пуассона.
Це припущення є певним наближенням до реальності. Моделювання вхідної  
стимуляції стохастичним процесом відображає той факт, що реальні 
послідовності вхідних стимулів мають виключно нерегулярний характер,
а їх статистика точно не означена.
Процес Пуассона береться, як найпростіший стохастичний процес.

Процесс Пуассона можна визначити як множину можливих траєкторій $n(t)$,
де $n(t)$ позначає число вхідних імпульсів, одержаних до моменту $t$, з
відповідним полем ймовірності на ній \cite[гл. ІІ, \S\ 9]{1956}. Для наших 
цілей більше підходить еквівалентний опис.

В цьому описі можливі траєкторії процесу Пуассона подаються через
послідовні часові моменти $\{l_1,l_2,\dots,l_k,\dots\}$ одержання вхідних
імпульсів. Зрозуміло, що
\begin{equation}\label{poslidovni}
0<l_1<l_2<\dots<l_k<\dots\,.
\end{equation}

Пуассонівська міра на циліндричних множинах траєкторій, означених через 
часові моменти одержання подій, має наступний вигляд:
\begin{equation}\label{PM2}
e^{-\la l_1}\la dl_1 e^{-\la(l_2- l_1)}\la dl_2\dots e^{-\la(l_k- l_{k-1})}\la dl_k,
\end{equation}
де $\la$ --- інтенсивність процесу Пуассона.
Ця міра задовольняє умови узгодженності (для перевірки слід врахувати (\ref{poslidovni})),
 отже єдиним чином продовжується до $\sigma$-аддитивної міри,
  узгодженої з тіхонівською в $\mathbb{R^N}$ топологією, \cite{Kolmogorov1974}.

\subsection{Вихідний потік}

Коли ІНВ одержує вхідні імпульси,  то в деякі
моменти часу відбуваються постріли, тобто нейрон надсилає вихідний
імпульс (спайк). Наша задача --- охарактеризувати вихідний потік імпульсів.

Зауважимо, що після кожного пострілу нейрон опиняється в стандартному
стані з $V=0$. Стан вхідного потоку (процесу Пуассона) незмінний в часі.
Отже, вихідний потік буде процесом відновлення і для його вичерпної
характеристики досить знати щільність $P(t)$ ймовірності розподілу
міжімпульсних (міжспайкових) інтервалів, МСІ. Вираз $P(t)\,dt$ дає 
ймовірність одержати міжспайковий інтервал в межах $[t;t+dt[$. Цю 
ймовірність можна обчислювати як ймовірність того, що
перший постріл нейрона відбудеться через $t$ одиниць часу після 
початку активності вхідного процесу (вмикання). При цьому в момент 
вмикання нейрон знаходиться в стані спокою: $V(0)=0$.

Задача відшукання $P(t)$ належить до класу граничних
задач для випадкових процесів і зводится до обчислення
часів першого досягнення
рівня $V_0$. Задачі такого роду досліджуються в звязку з масовим 
обслуговуванням, оцінками ризиків та ін. При цьому використовується модель
складного пуассонівського процесу зі знесенням (СППЗ).  Зокрема, в 
\cite[стор. 125]{1975} модель СППЗ використано для опису
нейронної активності (див. також п.\ref{Dissc}, нижче). 
В контексті нейрофізики знесення означає присутність
постійного за величиною гальмівного струму, який частково компенсує дію
вхідних збуджувальних імпульсів. Струм, який виникає внаслідок електричних втрат,
не постійний, а пропорційний величині $V(l)$. Це унеможливлює застосування
моделі СППЗ в ситуації, означеній в (\ref{tau}) (наявність електричних
втрат).

\section{Існування $P(t)$, неперервність по $t$}

Одержання першого після вмикання пострілу в момент $t$ відбувається
якщо мають місце дві незалежних події. Друга подія --- це одержання
вхідного імпульсу в інтервалі $[t;t+dt[$. Перша подія полягає в тому, що
всі попередні імпульси розташовані в часі так, що не викликають пострілу,
але створюють на момент $t$ в нейроні таке збудження ($V(t)$), що
одержання наступного вхідного в цей момент викличе постріл. Позначимо 
ймовірність першої події $\widetilde P(t)$. Тоді шукана ймовірність має вигляд
$$
P(t)dt=\widetilde P(t)\la dt.
$$

\begin{definition}
Послідовність $k$ вхідних імпульсів, або часових моментів їх одержання
$\{l_1,l_2,\dots,l_k\}$
називається мовчазною $k$-послідовністю, якщо під час одержання
нейроном цих імпульсів пострілу не відбувається при одержанні будь-якого з них.
\end{definition}

Позначимо через $\mathbb P_{k,t}$, $k=1,2,\dots$, подію, яка полягає в 
тому, що перші $k$ вхідних імпульсів,  $\{l_1,l_2,\dots,l_k\}$, складають
мовчазну $k$-послідовність, а одержання наступного вхідного в момент
$l_{k+1}\in[t;t+dt[$ викличе постріл. Позначимо через $\widetilde P_k(t)$
ймовірність події $\mathbb P_{k,t}$.

\begin{theorem}
Ймовірність $\widetilde P_k(t)$ існує і неперервна по $t$.
\end{theorem}

\noindent
Доведення. Позначимо через $\mathcal V_{l_1\dots l_i}(x)$ наступну функцію
\begin{equation}\label{potent}
\mathcal V_{l_1\dots l_i}(x)=h\sum\limits_{j=1}^i e^{-(x-l_j)/\tau}.
\end{equation}
Якщо $\{l_1\dots l_i\}$ --- мовчазна послідовність вхідних часових моментів
і $x>l_i$, то функція $\mathcal V_{l_1\dots l_i}(x)$ дає величину збудження в
нейроні в момент $x$ при одержанні ним вхідних імпульсів в моменти
$\{l_1\dots l_i\}$. 

Для того, щоб послідовність $\{l_1,l_2,\dots,l_k\}$ була мовчазною
необхідно і досить виконання наступних умов
\begin{equation}\label{M}
\mathcal V_{l_1}(l_2)\le V_0-h,\quad
\mathcal V_{l_1 l_2}(l_3)\le V_0-h,\ \dots,\ 
\mathcal V_{l_1\dots l_{k-1}}(l_k)\le V_0-h.
\end{equation}
Для того, щоб імпульс в часовий момент $l_{k+1}\in[t;t+dt[$ 
викликав постріл необхідно і досить виконання наступних умов:
\begin{equation}\label{obmezh}
l_k < t,
\end{equation}
\begin{equation}\label{postril}
\mathcal V_{l_1\dots l_k}(t) > V_0-h.
\end{equation}
Множина реалізацій процесу Пуассона $M_{k,t}$, яка відповідає події $\mathbb P_{k,t}$,
задається в $\mathbb{R^N}$ умовами (\ref{poslidovni}), (\ref{M}), (\ref{obmezh}) і 
(\ref{postril}). Оскільки всі ці умови формулюються за допомогою
нерівностей, заданих неперервними з $\mathbb{R^N}$ в $\mathbb{R}^1$
функціями, то $M_{k,t}$ --- борелівська і $\mathbb P_{k,t}$ --- корректна
подія. Отже, $\widetilde P_k(t)$ існує.

Для доведення неперервності $\widetilde P_k(t)$ по $t$ слід оцінити різницю
\begin{equation}\label{riznytsja}
|\widetilde P_k(t+\Delta t)-\widetilde P_k(t)|.
\end{equation}
Остання різниця дорівнює різниці мір (\ref{PM2}) множин $M_{k,t}$ і $M_{k,t+\Delta t}$.
Ця різниця не перевищує міри більшої з двох множин: $M_{k,t+\Delta t}\backslash M_{k,t}$ і 
$M_{k,t}\backslash M_{k,t+\Delta t}$. 

Множина $M_{k,t+\Delta t}\backslash M_{k,t}$ задається
умовами (\ref{poslidovni}), (\ref{M}) і наступними умовами
$$
t\le l_k < t+\Delta t,\quad
\mathcal V_{l_1\dots l_k}(t+\Delta t) > V_0-h.
$$
Перша з цих умов ґарантує, що міра (\ref{PM2}) 
множини $M_{k,t+\Delta t}\backslash M_{k,t}$ має порядок 
$O(\Delta t)$.

Множина $M_{k,t}\backslash M_{k,t+\Delta t}$ задається
умовами (\ref{poslidovni}), (\ref{M}) і наступними умовами
$$
\mathcal V_{l_1\dots l_k}(t) > V_0-h,\quad
\mathcal V_{l_1\dots l_k}(t+\Delta t) \le V_0-h.
$$
Враховуючи (\ref{potent}), перпишемо останнє в наступному вигляді:
\begin{equation}\label{Mdt}
 V_0-h < \mathcal  V_{l_1\dots l_k}(t)\le e^{\Delta t/\tau}(V_0-h).
\end{equation}
Позначимо $M_{\Delta t}$ множину в $\mathbb{R^N}$ у точок якої перші $k$ координат
 $\{l_1,\dots,l_k\}$ такі, що задовольняють (\ref{Mdt}). $M_{\Delta t}$ корректна подія і 
$M_{\Delta t}\supset M_{k,t}\backslash M_{k,t+\Delta t}$. Отже, міра (\ref{PM2})
множини $M_{k,t}\backslash M_{k,t+\Delta t}$ не перевищує міри $M_{\Delta t}$.
Крім того $\Delta t_2 > \Delta t_1 \Rightarrow M_{\Delta t_2}\supset M_{\Delta t_1}$ і $\bigcap\limits_{\Delta t>0} M_{\Delta t} = \emptyset$.
З останнього слідує, що границя міри $M_{\Delta t}$ дорівнює нулю коли $\Delta t \to 0$
і те ж саме має місце для $M_{k,t}\backslash M_{k,t+\Delta t}$.
Отже, різниця (\ref{riznytsja}) прямує до нуля коли $\Delta t \to 0$
і неперервність $\widetilde P_k(t)$ доведено.

\begin{theorem}
Щільність ймовірності МСІ $P(t)$ існує і неперервна по $t$.
\end{theorem}

\noindent
Доведення. Очевидно що вхідний імпульс, який спричиняє перший постріл
нейрона, може мати номер $n=2,3,\dots$ і випадки з різними $n$ несумісні,
звідки слідує
\begin{equation}\label{Sum}
P(t)dt=\sum\limits_{k\ge1} \widetilde P_k(t) \la dt.
\end{equation}
При фіксованому $t$ кількість доданків в (\ref{Sum}) скінченна. Дійсно, для 
того щоб два ізольованих послідовних імпульси не викликали пострілу 
нейрона вони мають бути розділені проміжком не коротшим від $T_2$, де
$T_2$ знаходиться з умови 
$
h e^{-T_2/\tau} = V_0 - h ,
$ 
або
\begin{equation}\label{T2}
T_2 = \tau\ln\frac{h}{V_0-h}.
\end{equation}
Присутність додаткових імпульсів лише посилює цю вимогу. Таким чином для 
кожного $t$ існує таке $k_{max}$, що розмістити на інтервалі $]0;t[$ 
мовчазну $m$-послідовність з $m>k_{max}$ неможливо і $\widetilde P_m(t')=0$
при $t'\in\,]0;t[$. Отже, $P(t)$ корректно визначається суммою (\ref{Sum}).

Для доведення неперервності $P(t)$ слід точніше з’ясувати які доданки 
містить сумма (\ref{Sum}). Означимо з цією метою ще один часовий інтервал
$T_3$: 
$
V_0 e^{-T_3/\tau} = V_0 - h,
$ 
або
\begin{equation}\label{T3}
T_3 = \tau\ln\frac{V_0}{V_0-h}.
\end{equation}
Подамо можливі значення МСІ як об’єднання множин, що не 
перетинаються:
\begin{equation}\label{Theta}
]0;\infty[ = ]0;T_2] + \sum\limits_{m=3}^\infty ]\Theta_m;\Theta_{m+1}],
\end{equation}
де
$$
\Theta_m=T_2+(m-3)T_3,\quad m=3,4,\dots,\quad \Theta_2=0.
$$
$\Theta_m$ є мінімальною довжиною мовчазної $(m-1)$-послідовності, де
довжина послідовності $\{l_1,l_2,\dots,l_k\}$ визначається як $l_k-l_1$.
Дійсно, послідовність з
\begin{equation}\label{Naykor}
l_1=0,\quad l_2=\Theta_3,\ \dots,\ l_{m-1}=\Theta_m
\end{equation}
мовчазна, оскільки з означень $T_2$, $T_3$ слідує, що для послідовності (\ref{Naykor})
мають місце наступні рівності:
\begin{equation}\label{Mum}
\mathcal V_{l_1}(l_2)= V_0-h,\quad
\mathcal V_{l_1 l_2}(l_3)= V_0-h,\ \dots,\ 
\mathcal V_{l_1\dots l_{m-2}}(l_{m-1})= V_0-h.
\end{equation}
Зменшення часової відстані між будь-якими двома сусідніми імпульсами
приведе до порушення одної з рівностей (\ref{Mum}) 
з заміною ``$=V_0-h$'' на ``$>V_0-h$'' і до пострілу в момент одержання 
другого з них. Сказане дозволяє переписати (\ref{Sum}) наступним 
чином
\begin{equation}\label{Sumi}
P(t)dt=\sum\limits_{k=1}^{m-1} \widetilde P_k(t) \la dt,\quad
t\in\, ]\Theta_m;\Theta_{m+1}],\quad m=2,3,\dots,
\end{equation}
звідки слідує неперервність $P(t)$ на інтервалах $]\Theta_m;\Theta_{m+1}[$
як сумми скінченного числа неперервних функцій. Для завершення слід 
довести неперервність $P(t)$ в точках $\Theta_{m+1}$, $m=2,3,\dots$.
Для цього досить довести рівність
\begin{equation}\label{nulj}
\widetilde P_m(\Theta_{m+1})=0,\quad m=2,3,\dots .
\end{equation}
Останнє випливає з того, що $\Theta_{m+1}$ --- найменша довжина
мовчазної $m$-послідовності. Найкоротша $m$-послідовність, яка
може бути розміщена на відрізку $[0;\Theta_{m+1}]$ єдина. Вона має
часові моменти означені в (\ref{Naykor}) і $l_m=\Theta_{m+1}$. Отже, 
циліндрична множина $M_{m,\Theta_{m+1}}$ має в основі циліндра
одну точку з $\mathbb{R}^m$ і
ймовірність події $\mathbb P_{m,\Theta_{m+1}}$ дорівнює нулю,
що доводить (\ref{nulj}).

\section{Окремі доданки в (\ref{Sumi})}

Для $k=2,3,\dots$ введемо наступні позначення

\begin{description}
\item[$P_k^0(t)\la dt$] --- ймовірність отримати $k$-послідовність $\{l_1,\dots,l_{k-1},l_k\in [t;t+dt[\}$ таку, що $\{l_1,\dots,l_{k-1}\}$ --- мовчазна.

\item[$P_k^-(t)\la dt$] --- ймовірність отримати мовчазну 
$\{l_1,\dots,l_{k-1},l_k\in [t;t+dt[\}$.
\end{description}

З означень слідує, що 
$$%\begin{equation}\label{nuli}
%\begin{aligned}
P_k^0(t) = 0,\quad \text{ якщо }\quad t\in]0;\Theta_k],
\qquad
P_k^-(t) = 0,\quad \text{ якщо }\quad t\in]0;\Theta_{k+1}],
%\end{aligned}
$$%\end{equation}
а також
$$
\widetilde P_k(t) \la dt = P_{k+1}^0(t)\la dt - P_{k+1}^-(t)\la dt.
$$

Використавши останнє перепишемо (\ref{Sumi}) в наступному вигляді
\begin{equation}\label{Sumiii}
P(t)dt= 
\sum\limits_{k=2}^{m-1} 
\left(P_k^0(t)\la dt - P_k^-(t)\la dt\right)
+P_m^0(t)\la dt,\,
t\in\,]\Theta_m;\Theta_{m+1}],\, m\ge2 .
\end{equation}

Зокрема, 
$
P(t)dt = P_2^0(t)\la dt\,\text{ якщо } t\in\,]\Theta_2;\Theta_3],
$
де
\begin{equation}\label{P20}
P_2^0(t)\la dt=\la t e^{-\la t}\la dt,
\end{equation}
Зауважимо, що якщо для $s>\Theta_{k+1}$ вираз
$P_k^-(s)\la ds$ дає ймовірність
одержати мовчазну $k$-послідовність 
$\{l_1,\dots,l_{k-1},l_k\in [s;s+ds[\}$, то вираз 
$P_k^-(s)\la ds\,e^{-\la(t-s)} \la\,dt$ для $t>s$
дає ймовірність одержати $(k+1)$-послідов\-ність 
$\{l_1, \dots, l_{k-1}, l_k \in [s;s+ds[, l_{k+1}\in [t;t+dt[\}$,
таку, що підпослідовність її перших $k$ часових моментів мовчазна.
З сказаного слідує наступне
\begin{equation}\label{F-la}
\int\limits_{\Theta_{k+1}}^t P_k^-(s)\la\, ds\,e^{-\la(t-s)}\la\,dt
= P_{k+1}^0(t)\la\,dt,\quad t\ge \Theta_{k+1},\quad
k=2,3,\dots .
\end{equation}
Отже, знаходження явного вигляду доданків в суммі (\ref{Sumiii})
зводиться до знаходження явних виразів для функцій 
$P_k^-(t),\, k=2,3,\dots.$ Нижче ми вказуємо явні вирази для цих функцій
у вигляді кратних інтегралів. Обчислення самих інтегралів виконано для
$k=2,3$. 	

Як вже відмічалось вище, $P_k^-(t)=0$ при $t\in\,]0;\Theta_{k+1}]$ 
і при $t=\Theta_{k+1}$ існує точно одна мовчазна $k$-послідовність
типу $\{l_1,\dots,l_{k-1},l_k = t\}$,
а саме $\{\Theta_2,\Theta_3,\dots,\Theta_{k+1}\}$. Для $t>\Theta_{k+1}$
ймовірність одержати мовчазну $k$-послідовність типу $\{l_1,\dots,l_{k-1},l_k\in [t;t+dt[\}$
 строго позитивна. Для її обчислення слід проінтегрувати вираз
$$
e^{-\la l_1}\la dl_1 e^{-\la(l_2-l_1)}\la dl_2\,\dots\, e^{-\la(t-l_{k-1})}\la dt 
$$
по множині значень координат $l_1,l_2,\dots,l_{k-1}$ таких, що
забезпечують відсутність пострілів при одержанні імпульсів в моменти
$l_1,l_2,\dots,l_{k-1},t$:
\begin{equation}\label{Pkmint}
P_k^-(t)\la dt = e^{-\la t}\la^k dt 
\int\limits_{\lbot1}^{\ltop1} dl_1
\int\limits_{\lbot2}^{\ltop2} dl_2
\dots
\int\limits_{\lbot{k-1}}^{\ltop{k-1}} dl_{k-1},
\end{equation}
де верхні і нижні границі інтегрування слід визначити.
Нижні границі визначаються з умов, що при одержанні імпульсів в 
моменти $l_1,l_2,\dots,l_{k-1},t$ не має відбутись пострілу.
Очевидно, $\lbot1=0$.
В загальному випадку $\lbot{i+1}(l_1,\dots,l_i)$ визначається
з умови 
$\mathcal V_{l_1\dots l_i}(\lbot{i+1})=V_0-h,$ звідки
\begin{equation}\label{lboti1}
\lbot{i+1}(l_1,\dots,l_i)= T_2 +
\tau\ln\left(\sum\limits_{j=1}^i e^{l_j/\tau}\right),
\quad i=1,\dots,k-2.
\end{equation}

Верхні межі інтегрування в (\ref{Pkmint}) залежать додатково від
значень $k$ і $t$: $\ltop{i+1}=\ltop{i+1}(k,t,l_1,\dots,l_i)$.
При цьому $\ltop{i+1}(k,t,l_1,\dots,l_i)$ має бути вибрана так,
щоб забезпечити можливість розміщення моментів часу 
$l_{i+2},\dots,l_{k-1},t$ так, що результуюча $k$-послідовність
$\{l_1,\dots,\ltop{i+1},\dots,l_{k-1},t\}$ не дає пострілів.

Для визначення $\ltop{i+1}(k,t,l_1,\dots,l_i)$ при вже зафіксованих
$l_1,\dots,l_i$ і $t$ слід зауважити, що з умови відсутності
пострілів слідує, що при виборі найбільших можливих значень для
$\ltop{i+1},l_{i+2},\dots,l_{k-1}$ виконуються наступні умови:
\begin{align}\label{ltopspiv}
\mathcal{V}_{l_1\dots l_i\, \ltop{i+1}}\,(l_{i+2}) \,=\,& V_0-h,
\\\nonumber
\dots&
\\\nonumber
\mathcal{V}_{l_1\dots l_i\,\ltop{i+1}\,l_{i+2}\dots l_{k-1}}(t) \,=\, &V_0-h.
\end{align}

З останнього слідує, що при виборі найбільших можливих значень для 
$\ltop{i+1},l_{i+2},\dots,l_{k-1}$ виконуються наступні умови:
$$
l_{i+3}-l_{i+2} = l_{i+4}-l_{i+3}= \,\dots\, =t-l_{k-1}=T_3.
$$
Отже, найбільше можливе значення для $l_{i+2}$
$$
l^r_{i+2}=t-(k-i-2)T_3.
$$
$\ltop{i+1}(k,t,l_1,\dots,l_i)$ тепер можна знайти підставивши 
$l^r_{i+2}$ замість $l_{i+2}$ в (\ref{ltopspiv}):
\begin{equation}\label{ltopi1}
\ltop{i+1}(k,t,l_1,\dots,l_i)=
\tau\ln\left(e^{(t-\Theta_{k+1-i})/\tau}-\sum\limits_{j=1}^i e^{l_j/\tau}\right),\quad
i=0,\dots,k-2.
\end{equation}

Формула (\ref{Pkmint}) разом з формулами (\ref{lboti1}), 
(\ref{ltopi1}), які в явному вигляді задають межі інтегрування в (\ref{Pkmint}),
дає явний вигляд для ймовірностей $P_k^-(t)\la dt$, $k=2,\dots$.

\subsection{Обчислення перших доданків в (\ref{Sumiii})}
Функцію розподілу $P(t)dt$ на перших ділянках розбиття (\ref{Theta})
тепер можна знайти в явному вигляді.
Обчислення за формулою (\ref{Pkmint}) при $k=2$, або $k=3$ дають
\begin{align}\label{P2m}
P_2^-(t) =\,& e^{-\lt}\la (t-T_2),\quad t\ge\Theta_3,
\\\label{P3m}
P_3^-(t) =\, & e^{-\lt}\la^2 \left((t-2T_2)(t-\Theta_4)-\frac{1}{2}(t-\Theta_4)^2\right)+
\\\nonumber
+\, &e^{-\lt}(\tau\la)^2
\left(\mathrm{Li}_2\left(e^{(T_2-t)/\tau}\right)-\mathrm{Li}_2\left(e^{-T_3/\tau}\right)\right),
\quad t\ge \Theta_4,
\end{align}
де $\mathrm{Li}_2$ позначає ділогарифм.

На основі формул (\ref{P2m}), (\ref{P3m}) і (\ref{F-la}) можна знайти
решту доданків в (\ref{Sumiii}) для значень $t\le\Theta_5$:
\begin{align}\label{P30}
P_3^0(t)=\,& e^{-\lt}\frac{\la^2(t-T_2)^2}{2},\quad t\ge\Theta_3,
\\\label{P40}
P_4^0(t)=\,& e^{-\la t}\frac{\la^3}{6}(\Theta_4 -t)^2(2T_3-4T_2+t)+
\\\nonumber
+\,&e^{-\la t}\tau^2\la^3 (\Theta_4-t)\mathrm{Li}_2\left(e^{-T_3/\tau}\right)+
\\\nonumber
+\,&e^{-\la t}(\tau\la)^3
\left(
\mathrm{Li}_3\left(e^{-T_3/\tau}\right)
-
\mathrm{Li}_3\left(e^{(T_2-t)/\tau}\right)
\right),\quad t\ge\Theta_4,
%Викладки в файлі:
% /home/alex/texts/texts.my/collabor/ukraine/Ksenia/papers/2012/lif-rozpodil/maxima/N02step3/P0.mac
\end{align}
де $\mathrm{Li}_3$ позначає трилогарифм.

\begin{figure}
\includegraphics[scale=0.35, angle=-90]{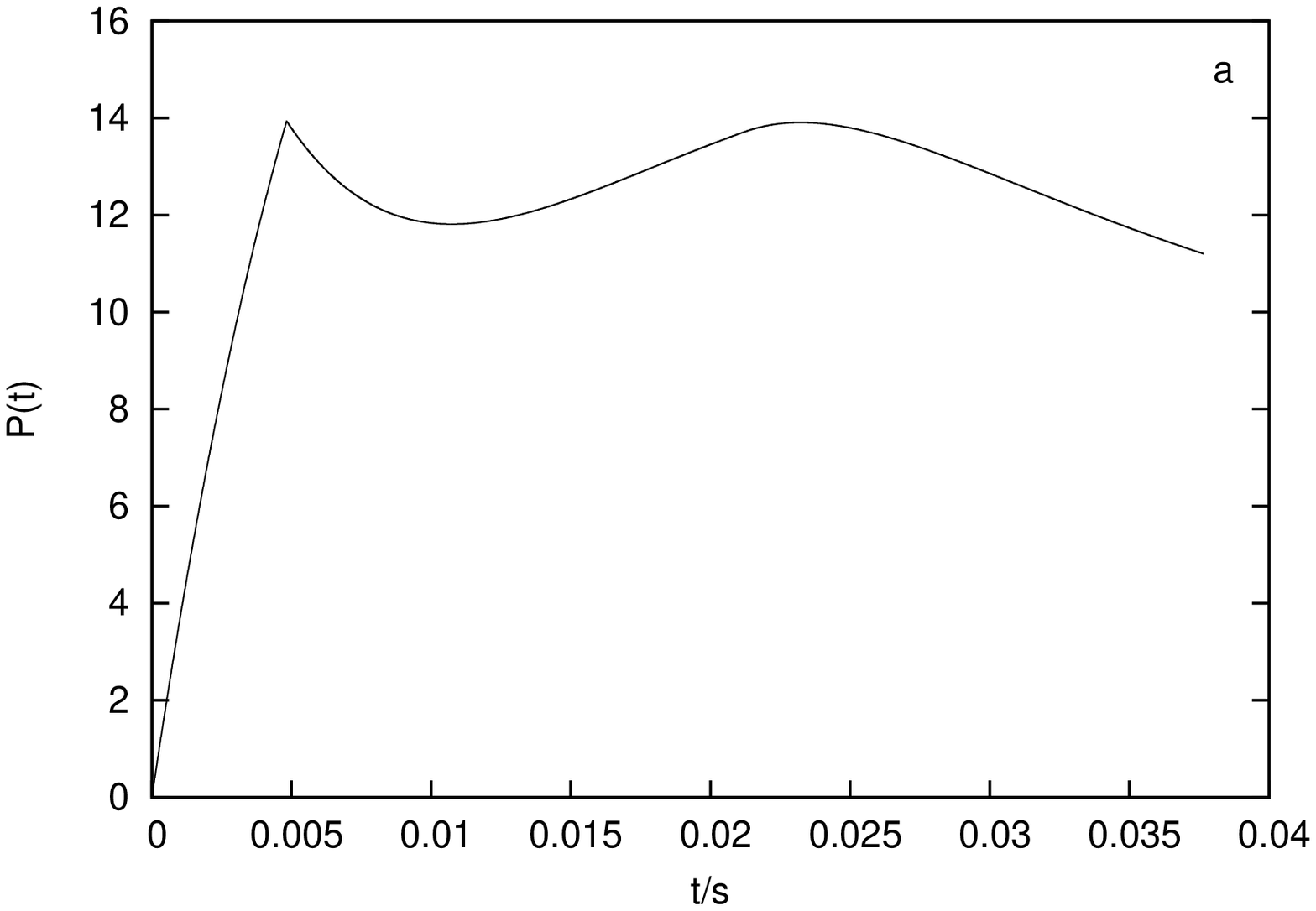}
\includegraphics[scale=0.35, angle=-90]{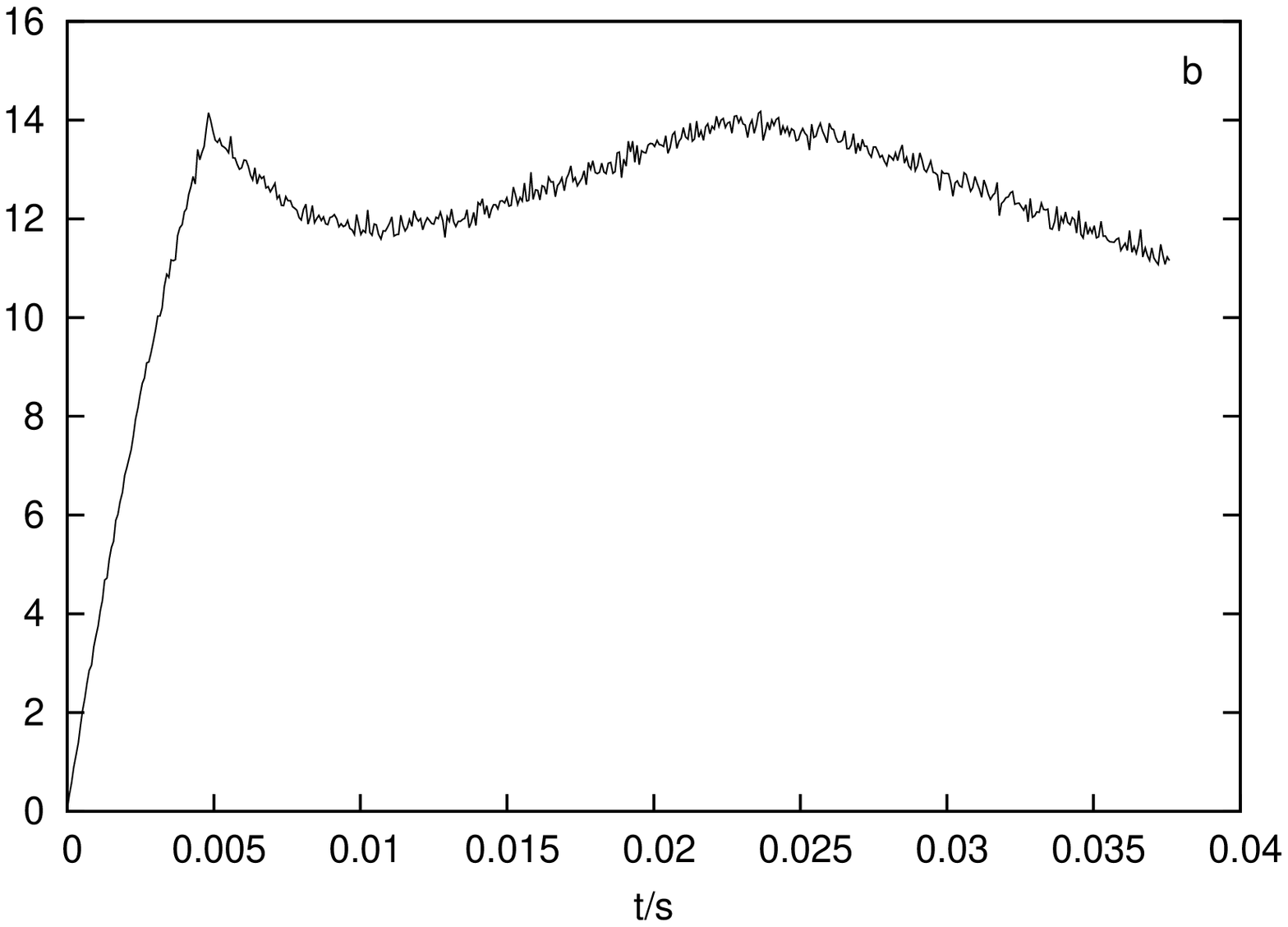}
\caption{\label{compare}Щільність ймовірності довжин МСІ, $P(t)$, (a) знайдена аналітично
за формулами (\ref{final}); 
(b) знайдена чисельно, симуляцією за методом Монте Карло. Тут
$\la =62.5 $ с$^{-1}$, $\tau=20$ мс, $V_0=20$ мВ, $h=11.2$ мВ.}
\end{figure}

Явні вирази (\ref{P20}), (\ref{P2m})--(\ref{P40}), дають явні вирази для 
формули (\ref{Sumiii}) на початковій ділянці значень $t$:
\begin{equation}\label{final}
P(t)dt =
\begin{cases}
\begin{alignedat}{2}
 &P_2^0(t)\la dt, &t\in\,]0;T_2],\\
 &\left(P_2^0(t)-P_2^-(t)+P_3^0(t)\right)\la dt, &t\in\,]T_2;T_2+T_3],\\
 &\left(\sum\limits_{k=2}^3\left(P_k^0(t)-P_k^-(t)\right)+P_4^0(t)\right)\la dt,\, &t\in]T_2+T_3;T_2+2T_3],
\end{alignedat}
\end{cases}
\end{equation}
де $T_2$, $T_3$ даються формулами (\ref{T2}), (\ref{T3}).
Графік виразу (\ref{final}) показано на Мал. \ref{compare}(a).

З метою перевірки одержаних виразів було розраховано чисельно
хід залежності $P(t)$ на відрізку $t\in ]0;\Theta_5]$ для певних значень параметрів $\lambda,\, 
\tau,\, V_0,$ $h$. Одержану залежність подано на Мал. \ref{compare}(b).

\section{Обговорення}\label{Dissc}

Одержаний розподіл має локальний мінімум на інтервалі $]\Theta_3;\Theta_5]$ при деяких значеннях
параметрів, Мал. \ref{compare}. Такого мінімуму не описано раніше для розрахунків в межах
диффузійного наближення, хоча мультимодальність в експериментальних гістограмах часто
спостерігається\footnote{Для нейронів включених в нейронну мережу головною причиною 
мультимодальності є затримані в часі зворотні зв’язки.}. 
При диффузійному наближенні, для генерації одного спайку потрібно одержати необмежену
кількість вхідних імпульсів, тоді як в даній роботі умова (\ref{cond}) забезпечує можливість
пострілу вже при дії двох близько розташованих в часі імпульсів. Це може пояснити 
мультимодальність Мал. \ref{compare}. Для математичних моделей, які комбінують диффузійний
процесс з випадковими стрибками скінченної величини, в чисельних розрахунках
 також отримано мультимодальність
густини розподілу ймовірностей, \cite[Fig. 5.4]{Sacerdote2013}.

Окремо слід відмітити роботу \cite{1967}, де розподіл МСІ обчислено без застосування
диффузійного наближення. В цій роботі замість (\ref{tau}) припускається, що одержаний 
імпульс зберігається в нейроні в незмінному вигляді протягом випадкового проміжку часу
після чого зникає. На перший погляд, це серйозна відмінність від детерміністичного 
експоненційого розпаду (формула (\ref{tau})),
який має місце в реальних нейронах. Але для збудження нейрона важливим є часовий хід
комплексного постсинаптичного потенціалу, який є суммою внесків від всіх одержаних імпульсів.
В припущеннях роботи \cite{1967}, реальний хід комплексного постсинаптичного потенціалу
апроксимується ступіньковою функцією. Така апроксимація тим точніша, чим більше імпульсів
необхідно для збудження нейрона до порогового значення. В \cite{1967} знайдені розподіли добре узгоджуються
з експериментальними коли МСІ не коротші від 60 мс. За цей час нейрон одержує значну кількість
вхідних імпульсів, що пояснює одержаний в \cite{1967} добрий збіг з експериментальними гістограмами.

Якщо порівняти одержані тут формули (\ref{Sumiii}), (\ref{Pkmint}) з 
формулами (2), (3) роботи \cite{Vidybida2007} 
для зв’язуючого нейрона (ЗН) з порогом 2, то можна помітити їх структурну подібність.
Це пояснюється тим, що умова (\ref{cond}) забезпечує поведінку ІНВ подібну до ЗН з порогом 2 в контексті підрахунку внесків окремих подій в ймовірність порогового збудження.

%\bibliographystyle{unsrt}

%\bibliography{References}{}

\end{document}